\documentclass{emulateapj}

\begin{document}

\shortauthors{Luhman \& Mamajek}
\shorttitle{Brown Dwarfs in Taurus}

\title{Spectroscopy of Putative Brown Dwarfs in Taurus}

\author{
K. L. Luhman\altaffilmark{1,2,3} and
E. E. Mamajek\altaffilmark{4}
}

\altaffiltext{1}{Department of Astronomy and Astrophysics, The Pennsylvania
State University, University Park, PA 16802; kluhman@astro.psu.edu.}
\altaffiltext{2}{Visiting Astronomer at the Infrared Telescope Facility, which
is operated by the University of Hawaii under Cooperative Agreement no.\ NCC
5-538 with the National Aeronautics and Space Administration (NASA),
Office of Space Science, Planetary Astronomy Program.}
\altaffiltext{3}{Center for Exoplanets and Habitable Worlds, The 
Pennsylvania State University, University Park, PA 16802.}
\altaffiltext{4}{Department of Physics and Astronomy,
The University of Rochester, Rochester, NY 14627.}

\begin{abstract}

Quanz and coworkers have reported the discovery of the coolest
known member of the Taurus star-forming complex (L2$\pm0.5$)
and Barrado and coworkers have identified a possible protostellar
binary brown dwarf in the same region.
We have performed infrared spectroscopy on the former and the brighter
component of the latter to verify their substellar nature.
The resulting spectra do not exhibit the strong steam absorption bands
that are expected for cool objects, demonstrating that they are not
young brown dwarfs.
The optical magnitudes and colors for these sources are also
indicative of background stars rather than members of Taurus.
Although the fainter component of the candidate protostellar binary
lacks spectroscopy, we conclude that it is a galaxy rather than
a substellar member of Taurus based on its colors and the constraints
on its proper motion.

\end{abstract}

\keywords{
stars: formation ---
brown dwarfs ---
binaries: visual ---
stars: pre-main sequence}

\section{Introduction}
\label{sec:intro}

Current surveys for brown dwarfs are exploring
new regimes of temperature, mass, metallicity, age, binarity, and
formation environment in order to address a variety of questions regarding
objects at low masses and cold temperatures.
For instance, brown dwarfs at low masses constrain the minimum mass of
the initial mass function while the youngest brown dwarfs in the protostellar
phase offer the most direct insight into the formation process of substellar
objects.

The Taurus star-forming region ($\tau\sim1$~Myr, $d=140$~pc) has been the 
target of two recent surveys for low-mass brown dwarfs and protostellar brown
dwarfs. \citet{qua10} searched for the former by obtaining near-infrared
(IR) images of a large sample of molecular cores in Taurus
and combining these data with optical photometry from the
Sloan Digital Sky Survey \cite[SDSS;][]{yor00,fin04}.
In the overlapping area of 1~deg$^{2}$ between the IR and optical images, 
\citet{qua10} identified five new brown dwarf candidates and performed
spectroscopy on two of them to assess their nature.
Using these spectra, they classified one of the candidates, CAHA~Tau~1,
as a young brown dwarf with a spectral type of L2, which would make it the
coolest (and likely least massive) known member of Taurus \citep{luh09tau}.

In another survey of Taurus, \citet{bar09} used archival mid-IR images from
the {\it Spitzer Space Telescope} \citep{wer04} to search for protostellar
brown dwarfs. They reported the discovery of a $2\farcs5$ pair of candidates
that could comprise a binary system, which were designated as
SSTB213 J041757.75+274105.5 A and B (henceforth
J041757~A and B). Although spectroscopy was unavailable for these objects,
\citet{bar09} concluded that the proper motion of the A component supports
its membership in Taurus and that the colors of the B component are
inconsistent with a galaxy, which is the primary source of contamination
in a survey for protostellar brown dwarfs.

We have performed IR spectroscopy on CAHA~Tau~1
and J041757~A to determine whether they are young brown dwarfs. 
We also have used the available colors and proper motion measurements
for these two objects and the remaining candidates from
\citet{qua10} and \citet{bar09} to determine if they are likely to be
substellar members of Taurus.

\section{Observations}

\begin{figure}
\plotone{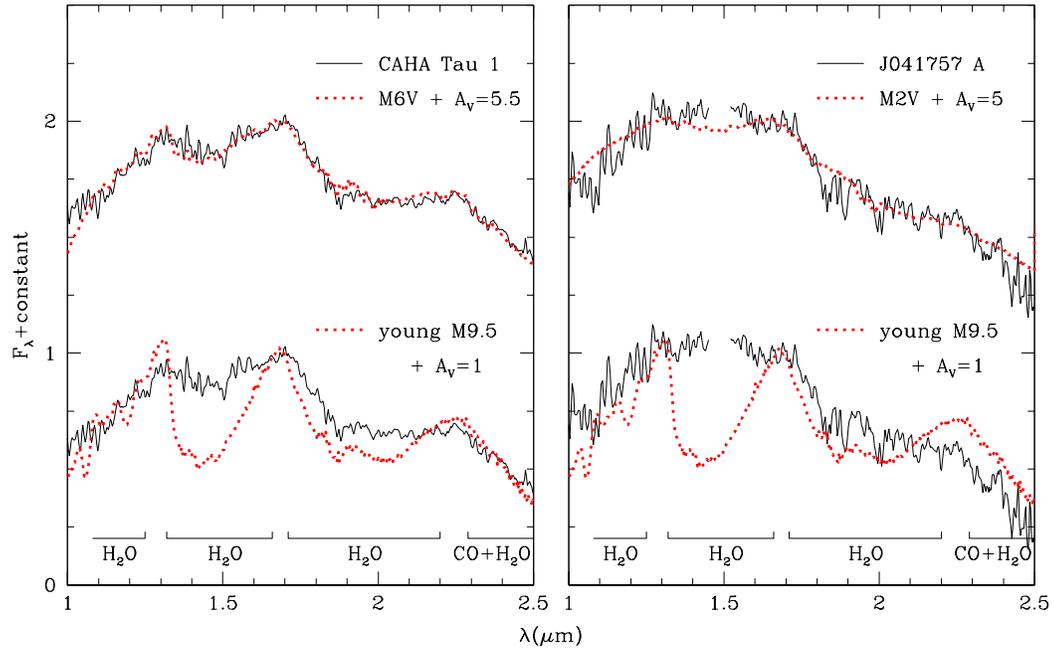}
\caption{
SpeX near-IR spectra of the brown dwarf candidates CAHA~Tau~1 and J041757~A
(solid lines) compared to data for dwarf and pre-main-sequence standards
(dotted lines).
These objects do not exhibit the strong steam absorption
bands that are expected for young late-type objects, as illustrated in this
comparison to the Taurus member KPNO~4 \citep[M9.5;][]{bri02}.
Instead, CAHA~Tau~1 agrees well with a reddened M6 dwarf (Gl~406) and
J041757~A is earlier than $\sim$M2 based on its lack of steam absorption.
These data have a resolution of $R\sim135$ and are normalized at 1.68~\micron.
}
\label{fig:spec}
\end{figure}

We obtained low-resolution near-IR spectra of CAHA~Tau~1 
and J041757~A with SpeX \citep{ray03} at the NASA
Infrared Telescope Facility (IRTF). We used the prism mode of SpeX with a
$0\farcs8$ slit, which provided a wavelength coverage of 0.8-2.5~\micron\ and
a resolution of $R\sim150$. We collected 12 and 16 exposures of CAHA~Tau~1 on
the nights of 2010 January 3 and 4, respectively, and 20 exposures of
J041757~A on the night of 2010 January 16.
The images had exposure times of 90~s and were taken during dither sequences
between two positions along the slit.
The observations of CAHA~Tau~1 were performed with the slit rotated to the
parallactic angle. For J041757~A, we aligned the slit so that it encompassed
the A and B sources simultaneously. Spectra that are not obtained at the
parallactic angle are susceptible to wavelength-dependent slit losses, but
such losses should be negligible for the observations of J041757 since they
were performed at a very low airmass ($\leq1.03$). 
These data were reduced with the Spextool
package \citep{cus04} and corrected for telluric absorption \citep{vac03}.
The average signal-to-noise ratios were $\sim30$ and 12 per pixel
for CAHA~Tau~1 and J041757~A, respectively.
We smoothed the reduced spectra to a slightly lower resolution ($R\sim135$)
to improve these ratios. The resulting spectra for CAHA~Tau~1
and J041757~A are presented in Figure~\ref{fig:spec}. A small portion of the
spectrum of J041757~A is not shown because it falls between the atmospheric
windows where the telluric correction was very poor.
The detection of J041757~B was too weak for a useful spectrum.

\section{Analysis}

\subsection{Spectral Classification}
\label{sec:class}

Using IR spectroscopy, \citet{qua10} classified CAHA~Tau~1 as a
young brown dwarf with a spectral type of L2$\pm$0.5.
If J041757~A is a member of Taurus, \citet{bar09} estimated 
that it should have a temperature of 1550--1750~K based on a comparison
of its photometry to the fluxes predicted by theoretical evolutionary models,
which also corresponds to an L spectral type \citep{dah02}.
To assess the accuracy of these classifications, we begin by comparing our
spectra of CAHA~Tau~1 and J041757~A in Figure~\ref{fig:spec} to a spectrum
of one of the coolest known members of Taurus, KPNO~4 \citep[M9.5;][]{bri02},
which has been reddened to roughly match the spectral slopes of the two
candidates. The most prominent features in the spectrum of KPNO~4 are the deep
H$_2$O absorption bands. CAHA~Tau~1 and J041757~A should show H$_2$O bands that
are as strong or stronger as those in KPNO~4 if they are L-type members
of Taurus, but this is not the case.
Instead, CAHA~Tau~1 exhibits weak H$_2$O absorption that is indicative of
a spectral type of M5--M6 if it is a dwarf and M3--M4 if it is a young
star\footnote{H$_2$O absorption is stronger in pre-main-sequence objects than in
dwarfs at a given optical spectral type \citep{lr99,luc01,mc04}.}.
The absence of detectable H$_2$O absorption in the spectrum of J041757~A
places constraints of $\lesssim$M2 (dwarf) and $\lesssim$M0 (young) on
the spectral type. 
The near-IR absorption features from the photospheres of young stars can be
diluted by continuum veiling from circumstellar dust emission, but this
is not a possible explanation for the absence of strong H$_2$O absorption
for CAHA~Tau~1 and J041757~A since they do not exhibit IR excess
emission (Section~\ref{sec:mid}).
Both candidates are much too faint to be members of Taurus with spectral
types of M3--M4 and $\lesssim$M0. Thus, we classify them as background stars
rather than substellar members of Taurus.

\subsection{Photometric Classification}

\subsubsection{Sources of Photometry}

In addition to spectroscopy, we have used the
available photometry to examine whether the candidates from
\citet{qua10} and \citet{bar09} are likely to be young brown dwarfs.
We have retrieved photometry in the five optical bands of SDSS
\citep[$ugriz$;][]{fuk96} from the Sixth Data Release of the survey
\citep{ade08} for CAHA~Tau~1--5 and J041757~A
(the B component was not detected by SDSS).
The calibration of these images is described by \citet{pad08}.
We selected the data measured with an aperture radius of $1.745\arcsec$.
\citet{bar09} measured $i$ and $z$ photometry for J041757~A and B
from archival images from the Canada-France-Hawaii Telescope (CFHT).
The data for J041757~A from \citet{bar09} agree with those
from SDSS, which suggests that the two photometric systems are similar.
Therefore, we have adopted the $i$ and $z$ photometry from \citet{bar09}
for J041757~B.

\begin{figure}
\plotone{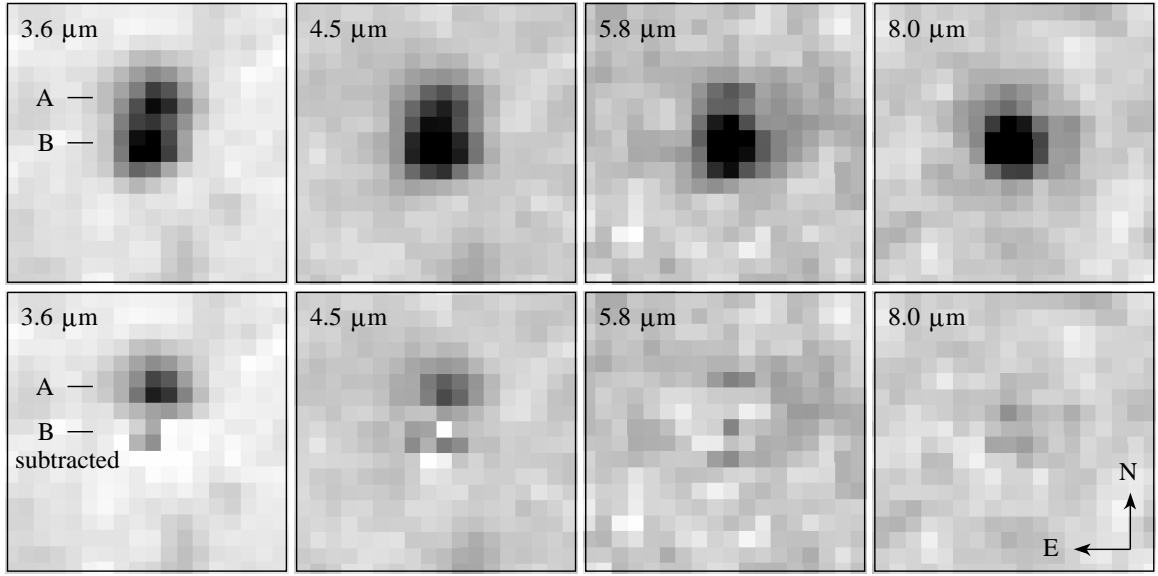}
\caption{
{\it Spitzer} images of the brown dwarf candidates J041757~A and B before
and after PSF subtraction of the B component. The images have a size
of $15\arcsec\times15\arcsec$ and are displayed on a logarithmic scale.
}
\label{fig:sub}
\end{figure}

We have compiled photometry at $J$, $H$, and $K_s$ from the Two-Micron All-Sky
Survey \citep[2MASS;][]{skr06} for CAHA~Tau~1, 2, 4, and 5, from
\citet{qua10} for CAHA~Tau~3, and from \citet{bar09} for J041757~A and B.
All of these sources also appear within archival images at 3.6, 4.5, 5.8,
8.0, and 24~\micron\ that were obtained by the {\it Spitzer Space Telescope}.
We have measured photometry from these images using the methods
that were applied to the known members of Taurus by \citet{luh10tau}.
J041757~B becomes increasingly dominant over J041757~A at longer wavelengths,
as shown in Figure~\ref{fig:sub}. To measure photometry at 3.6 and
4.5~\micron\ for J041757~A, we subtracted a scaled point spread function (PSF)
at the location of J041757~B \citep{mar06}. The A component was not detected in
the PSF-subtracted images at 5.8 and 8.0~\micron.
The spatial resolution of the image at 24~\micron\ is too low for resolving
J041757~A and B. We have assigned all of the 24~\micron\ flux to J041757~B
since it is much redder than J041757~A at 3.6--8.0~\micron.
We have adopted the average measurement in a given band if an object
was observed at multiple epochs by {\it Spitzer}.
We did not measure photometry from the second epoch of images at
3.6~\micron\ for J041757~A and B because they were contaminated by cosmic rays.
Our measurements of {\it Spitzer} photometry for CAHA~Tau~1--5 and J041757~A
and B are presented in Table~\ref{tab:phot}. 

\subsubsection{Optical and Near-IR Colors}

\begin{figure}
\plotone{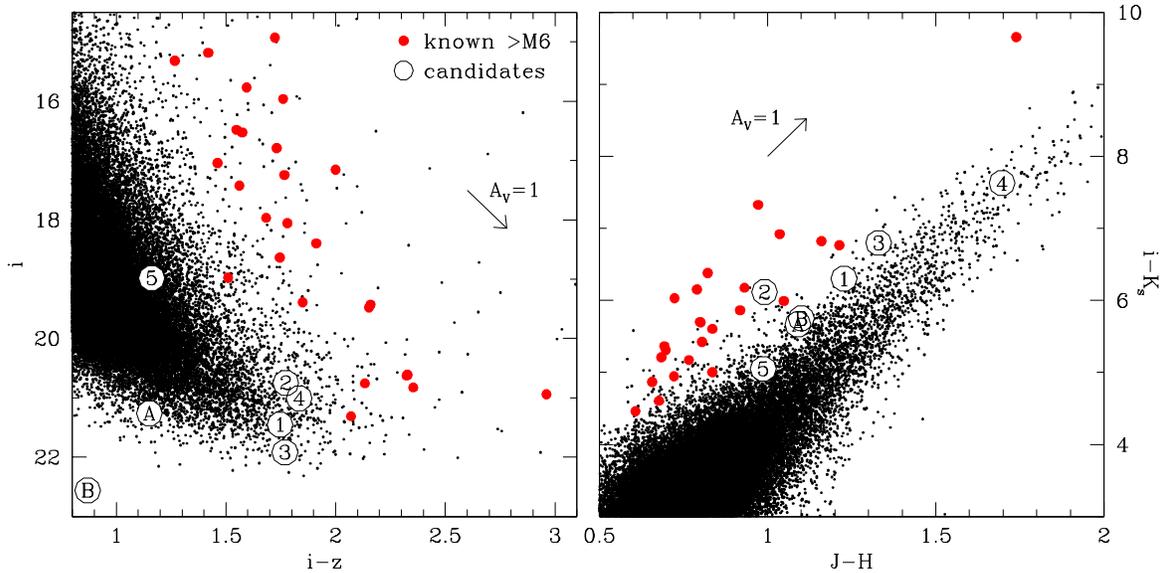}
\caption{
Optical and near-IR color-magnitude and color-color diagrams for the brown
dwarf candidates J041757~A and B and CAHA~Tau~1--5 
\citep[circles,][]{bar09,qua10},
known late-type members of Taurus ($>$M6, large points), and other sources
detected in the SDSS survey of Taurus \citep[small points,][]{fin04}.
The candidates appear below the sequence of known Taurus members in 
$i$ vs. $i-z$, indicating that they are probably field stars or
galaxies. One of the candidates, CAHA~Tau~2, has $J-H$ and $i-K_s$ colors
that are consistent with a spectral type later than M6, but
it is probably a field dwarf rather than a member of Taurus
given its position in $i$ vs. $i-z$.
The $i$ and $z$ data are from SDSS for all sources except J041757~B,
which was measured with deeper archival CFHT images by \citet{bar09}.
}
\label{fig:col}
\end{figure}

To determine if the candidates from \citet{qua10} and \citet{bar09} have
the appropriate optical colors and magnitudes for low-mass members of Taurus,
we plot them on a diagram of $i$ versus $i-z$ in Figure~\ref{fig:col}
with all known late-type members of Taurus ($>$M6) that were detected by SDSS.
We include in that diagram all other point sources in the SDSS images
that have photometric uncertainties less than 0.1~mag.
All of the candidates appear below the sequence of known members, which
indicates that they are probably field stars or galaxies.
CAHA~Tau 2, 4, and 5 were detected in the survey by \citet{luh04tau} and
were identified as probable nonmembers for the same reason.
One additional candidate, CAHA~Tau~3, was encompassed by the images from
\citet{luh04tau}, but its signal-to-noise ratio was too low for useful
photometry.
Young stars that are occulted by circumstellar disks and seen in scattered
light can appear anomalously faint for their colors, but J041757~B is
the only candidate that shows possible evidence of a disk in its
mid-IR photometry (Section~\ref{sec:mid}).

A diagram of $i-K_s$ versus $J-H$ is useful for distinguishing between
late-type objects and reddened stars at earlier types \citep{luh00tau}.
We show a diagram of this kind in Figure~\ref{fig:col} for the
candidates from \citet{qua10} and \citet{bar09}, the known late-type
members of Taurus, and all other sources with errors less than 0.1~mag
in the SDSS and 2MASS images of Taurus.
CAHA~Tau~2 is the only candidate that has the appropriate colors for
a young object later than M6. Given its position in $i$ versus $i-z$, 
it is probably a field M dwarf rather than a low-mass member of Taurus.

\subsubsection{Mid-IR Colors}
\label{sec:mid}

Approximately half of the known brown dwarfs in Taurus exhibit red mid-IR colors
that indicate the presence of circumstellar disks \citep{luh10tau}.
We can check for this signature of youth in the mid-IR photometry for
the candidate brown dwarfs from \citet{qua10} and \citet{bar09}.
According to our {\it Spitzer} measurements in Table~\ref{tab:phot}, 
CAHA~Tau~1--5 and J041757~A have neutral colors ($<0.2$) that agree
with those of stellar photospheres, and thus do not show evidence of disks.
The remaining candidate, J041757~B, does have a very red IR spectral energy
distribution, which was the basis of its original identification as a
candidate protostellar brown dwarf by \citet{bar09}.
As demonstrated in that study, the mid-IR colors of J041757~B are similar
to those of known protostars in Taurus.

\begin{figure}
\plotone{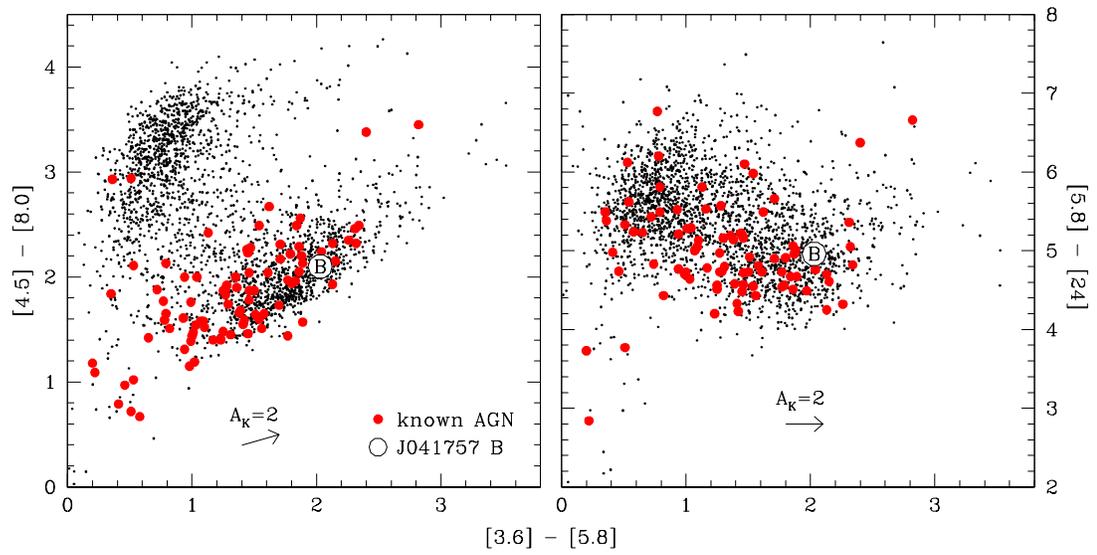}
\caption{
Mid-IR color-color diagrams for the brown dwarf candidate J041757~B
(circle, Table~1), spectroscopically confirmed AGNs from SWIRE 
\citep[large points,][]{sac09}, and faint sources ($[3.6]>11$) detected
in {\it Spitzer} images of Taurus \citep[small points,][]{luh10tau}.
The colors of J041757~B are consistent with those of an AGN.
}
\label{fig:irac}
\end{figure}

Based on a comparison to extragalactic surveys with {\it Spitzer}, 
\citet{bar09} concluded that a single type of galaxy could not account
for all of the mid-IR colors of J041757~B.
However, we find that the {\it Spitzer} colors of this object are fully
consistent with those of active galactic nuclei
\citep[AGN,][]{lacy04,lacy07,ste05,don07,bar08}.
This is illustrated in Figure~\ref{fig:irac},
where we show two {\it Spitzer} color-color diagrams for J041757~B 
and a sample of sources that have been spectroscopically classified as
AGNs \citep{sac09} from the {\it Spitzer} Wide-area Infrared Extragalactic
Survey \citep[SWIRE,][]{lon03,lon04}.
ELAISC15 J003603-433152 is an example of an AGN from SWIRE whose
{\it Spitzer} colors agree closely ($\leq0.16$~mag) with those of J041757~B.
We also include in Figure~\ref{fig:irac} all sources detected in
{\it Spitzer} images of Taurus that have $[3.6]>11$, photometric
errors less than 0.1~mag in all bands, and are not known members of
the star-forming region \citep{luh09tau,luh10tau}.
This sample contains $\sim2400$ objects and should be
dominated by background sources, mostly AGN and other red galaxies.
The mid-IR colors of J041757~B place it within this background population
as well.

After examining the colors of J041757~B at optical and near-IR wavelengths,
\citet{bar09} found that it is redder than galaxies from extragalactic
surveys like SWIRE in $I-J$ and $J-[3.6]$, leading them to conclude
that it is not a galaxy.
However, as shown in Figure~6 from \citet{bar09}, the colors of
J041757~B are consistent with a galaxy reddened by $A_V\sim5$, and 
the presence of extinction is naturally explained by the Taurus dark cloud
against which J041757~B is projected. Indeed, the shape of the spectrum of
J041757~A, which is a background star near the same line of sight, 
exhibits a reddening that corresponds to $A_V\gtrsim5$.
The dust map from \citet{sch98} indicates a similar extinction at
this position, although its spatial resolution is low
(FWHM=$6\farcm1$).

\subsection{Proper Motions}

\citet{qua10} and \citet{bar09} measured proper motions for their
candidates and compared these data to proper motions of known Taurus
members to assess whether they are likely to be members.
In this section, we examine the proper motion analysis from each of those
studies.

\citet{qua10} found that the proper motions of CAHA Tau~1 and 2 
agreed with those of the known members of Taurus. However, the motions
of the two candidates differed by only 1.3 and 1.9~$\sigma$, respectively,
from a simulated population of background giants because of the large
proper motion uncertainties. According to \citet{qua10},
the sizes of the motions of CAHA Tau~1 and 2 were also consistent with
those expected for field dwarfs.
Indeed, we classified CAHA Tau~1 as a M6 dwarf based on our spectroscopy
in Section~\ref{sec:class}.
\citet{qua10} concluded that the uncertainties in the proper motions 
for the remaining candidates, CAHA Tau~3--5, were too large for useful
constraints on their membership.

\citet{bar09} compared their proper motion measurements for J041757~A and B
to the mean and standard deviation of the motions of known Taurus
members that were computed by \citet{ber07}. Based on this comparison,
they concluded that J041757~A is a member of Taurus while the data for
J041757~B are inconclusive.
However, the proper motion data from \citet{ber07} are inadequate for
identifying likely members in this manner because they include
a large number of off-cloud stars that may be unrelated to Taurus,
which results in a proper motion dispersion that is erroneously large.
When a more reliable census of Taurus members is considered,
the mean proper motion and radial velocities of the Taurus subgroups 
indicate that the variation among the velocity vectors of the subgroups
is rather small \citep[$\pm$1--1.5~km\,s$^{-1}$,][]{luh09tau}.
Therefore, to assess the statistical likelihood of membership for
J041757~A and B, we compare their proper motions to the 
mean motion of the nearest Taurus subgroup (II):
$\mu_{\alpha}, \mu_{\delta} = +6.0\pm1$, $-26.8\pm1$ mas/yr
\citep{luh09tau}.
We assume that the absolute proper motions of J041757~A and B
are the same as the relative values measured by \citet{bar09}:
$\mu_{\alpha}, \mu_{\delta}$ = $-1.5\pm4.1$, $-20.5\pm3.5$ mas/yr for A
and $+2.0\pm4.0$, $-5.5\pm3.2$ mas/yr for B.
When we account for the proper motion errors and the uncertainty 
in the mean motion of the Taurus II subgroup, and we assume
a one-dimensional velocity dispersion of 1~km\,s$^{-1}$
(1.5 mas~yr$^{-1}$) for bona fide subgroup members, we
find that $\sim$7\% and 3$\times$10$^{-6}$\% of bona fide members
of Taurus II would have motions more deviant than that of 
J041757~A and B, respectively. Thus, we conclude that
the motion of J041757~A is just within $\sim$2$\sigma$ of the
mean motion of the subgroup, but the motion of J041757~B is
clearly inconsistent with membership.
Meanwhile, $\sim$20\% of objects with no proper motion 
should exhibit measurable proper motions larger than that of J041757~B.
Thus, the proper motion constraint for J041757~B is statistically consistent
with the zero motion expected for a galaxy.

\section{Conclusions}

We have used IR spectra, optical and IR colors, and proper motion measurements
to examine whether CAHA~Tau~1--5 \citep{qua10} and J041757~A and B 
\citep{bar09} are substellar members of the Taurus star-forming region.
We have obtained spectra of two of these sources, CAHA~Tau~1 and J041757~A.
The absence of strong steam absorption in their spectra 
indicates that they are not young brown dwarfs.
We classify CAHA~Tau~1 and J041757~A as reddened background stars with spectral
types of M5--M6 and $\lesssim$M2, respectively.
All of the candidates from \citet{qua10} and \citet{bar09}
appear below the sequence of known members of Taurus on an optical
color-magnitude diagram, indicating that they are field stars or galaxies.
One of these sources, CAHA~Tau~2, has $J-H$ and $I-K_s$ colors that are
consistent with a late spectral type ($>$M6), but
it is probably a field dwarf given its position in the optical color-magnitude
diagram. Six of the candidates do not show excess emission from disks in their
mid-IR colors. The remaining source, J041757~B, has a red spectral energy
distribution that is consistent with both a protostar and an AGN.
\citet{bar09} concluded that this object is too red in $I-J$ and $J-[3.6]$
to be a galaxy, but the reddening of these colors relative to galaxies
at high latitudes is easily explained by the Taurus dark cloud along its
line of sight.
The available proper motion measurements for CAHA~Tau~1--5 do not provide
strong constraints on their membership in Taurus.
We estimate a membership probability of 7\% for J041757~A
by comparing its proper motion to those of known Taurus members.
The motion of this object is consistent with a field dwarf, which is how we
classify it spectroscopically.
The proper motion constraints for J041757~B are inconsistent with
membership in Taurus and agree with the absence of motion expected for a
galaxy. Based on these various data, we conclude that CAHA~Tau~1--5
and J041757~A and B are not substellar members of Taurus.

\acknowledgements

K. L. was supported by grant AST-0544588 from the National Science
Foundation. We thank Pauline Barmby for helpful advice regarding the
IR colors of galaxies.

\clearpage

\begin{deluxetable}{llllll}
\tabletypesize{\scriptsize}
\tablewidth{0pt}
\tablecaption{{\it Spitzer} Photometry for Brown Dwarf Candidates in
Taurus\label{tab:phot}}
\tablehead{
\colhead{ID} &
\colhead{[3.6]} &
\colhead{[4.5]} &
\colhead{[5.8]} &
\colhead{[8.0]} &
\colhead{[24]} 
}
\startdata
CAHA Tau 1 & 14.63$\pm$0.03 & 14.42$\pm$0.03 & 14.5$\pm$0.1 & \nodata & \nodata \\
CAHA Tau 2 & 14.05$\pm$0.02 & 13.98$\pm$0.02 & 13.87$\pm$0.05 & \nodata & \nodata \\
CAHA Tau 3 & 14.53$\pm$0.04 & 14.40$\pm$0.05 & \nodata & \nodata & \nodata \\
CAHA Tau 4 & 12.87$\pm$0.02 & 12.75$\pm$0.02 & 12.66$\pm$0.03 & 12.69$\pm$0.03 & \nodata \\
CAHA Tau 5 & 13.59$\pm$0.02 & 13.46$\pm$0.02 & 13.42$\pm$0.03 & 13.40$\pm$0.04 & \nodata \\
J041757~A & 15.39$\pm$0.05 & 15.27$\pm$0.07 & $>$14.5 & $>$13.5 & \nodata \\
J041757~B & 14.81$\pm$0.02 & 13.80$\pm$0.02 & 12.78$\pm$0.03 & 11.70$\pm$0.03 & 7.82$\pm$0.04 \\
\enddata
\end{deluxetable}

\clearpage

\begin{thebibliography}{}

\bibitem[Adelman-McCarthy et al.(2008)]{ade08}
Adelman-McCarthy, J., et al. 2008, \apjs, 175, 297

\bibitem[Barmby et al.(2008)]{bar08}
Barmby, P., et al. 2008, \apjs, 177, 431

\bibitem[Barrado et al.(2009)]{bar09}
Barrado, D., et al. 2009, \aap, 508, 859

\bibitem[Bertout et al.(2007)]{ber07}
Bertout, C., Siess, L., \& Cabrit, S. 2007, \aap, 473, L21

\bibitem[Brice\~no et al.(2002)]{bri02}
Brice\~{n}o, C., Luhman, K. L., Hartmann, L., Stauffer, J. R., \& Kirkpatrick,
J. D. 2002, \apj, 580, 317

\bibitem[Cushing et al.(2004)]{cus04}
Cushing, M. C., Vacca, W. D., \& Rayner, J. T. 2004, \pasp, 116, 362

\bibitem[Dahn et al.(2002)]{dah02}
Dahn, C. C., et al. 2002, \aj, 124, 1170

\bibitem[Donley et al.(2007)]{don07}
Donley, J. L., Rieke, G. H., P\'{e}rez-Gonz\'{a}lez, P. G., Rigby, J. R., \&
Alonso-Herrero, A. 2007, \apj, 660, 167

\bibitem[Finkbeiner et al.(2004)]{fin04}
Finkbeiner, D. P., et al. 2004, \aj, 128, 2577

\bibitem[Fukugita et al.(1996)]{fuk96}
Fukugita, M., Ichikawa, T., Gunn, J. E., Doi, M., Shimasaku, K., \&
Schneider, D. P. 1996, \aj, 111, 1748

\bibitem[Lacy et al.(2004)]{lacy04}
Lacy, M., et al. 2004, \apjs, 154, 166

\bibitem[Lacy et al.(2007)]{lacy07}
Lacy, M., et al. 2007, \aj, 133, 186

\bibitem[Lonsdale et al.(2003)]{lon03}
Lonsdale, C. J., et al. 2003, \pasp, 115, 897

\bibitem[Lonsdale et al.(2004)]{lon04}
Lonsdale, C. J., et al. 2004, \apjs, 154, 54

\bibitem[Lucas et al.(2001)]{luc01}
Lucas, P. W., Roche, P. F., Allard, F., \& Hauschildt, P. H. 2001, \mnras,
326, 695

\bibitem[Luhman(2000)]{luh00tau}
Luhman, K. L. 2000, \apj, 544, 1044

\bibitem[Luhman(2004)]{luh04tau}
Luhman, K. L. 2004, \apj, 617, 1216

\bibitem[Luhman(2006)]{luh06}
Luhman, K. L. 2006, \apj, 645, 676

\bibitem[Luhman \& Rieke(1999)]{lr99}
Luhman, K. L., \& Rieke, G. H. 1999, \apj, 525, 440

\bibitem[Luhman et al.(2010)]{luh10tau}
Luhman, K. L., Allen, P. R., Espaillat, C., Hartmann, L., \& Calvet, N.
2010, \apjs, 186, 111


\bibitem[Luhman et al.(2009)]{luh09tau}
Luhman, K. L., Mamajek, E. E., Cruz, K. L., \& Allen, P. R. 2009, \apj, 703, 399

\bibitem[Marengo et al.(2006)]{mar06}
Marengo, M., Megeath, S. T., Fazio, G. G., Stapelfeldt, K. R., Werner, M. W.,
Backman, D. E. 2006, \apj, 647, 1437

\bibitem[McGovern et al.(2004)]{mc04}
McGovern, M. R., Kirkpatrick, J. D., McLean, I. S., Burgasser, A. J.,
Prato, L., \& Lowrance, P. J. 2004, \apj, 600, 1020

\bibitem[Padmanabhan et al.(2008)]{pad08}
Padmanabhan, N., et al. 2008, \apj, 674, 1217

\bibitem[Quanz et al.(2010)]{qua10}
Quanz, S. P., Goldman, B., Henning, T., Brandner, W., Burrows, A.,
\& Hofstetter, L. W. 2010, \apj, 708, 770

\bibitem[Rayner et al.(2003)]{ray03}
Rayner, J. T., et al. 2003, \pasp, 115, 362

\bibitem[Sacchi et al.(2009)]{sac09}
Sacchi, N., et al. 2009, \apj, 703, 1778

\bibitem[Schlegel et al.(1998)]{sch98}
Schlegel, D. J., Finkbeiner, D. P., \& Davis, M. 1998, \apj, 500, 525

\bibitem[Skrutskie et al.(2006)]{skr06}
Skrutskie, M., et al. 2006, \aj, 131, 1163

\bibitem[Stern et al.(2005)]{ste05}
Stern, D., et al. 2005, \apj, 631, 163

\bibitem[Vacca et al.(2003)]{vac03}
Vacca, W. D., Cushing, M. C., \& Rayner J. T., 2003, \pasp, 115, 389

\bibitem[Werner et al.(2004)]{wer04}
Werner, M. W., et al. 2004, \apjs, 154, 1

\bibitem[York et al.(2000)]{yor00}
York, D. G., et al. 2000, \aj, 120, 1579

\end{thebibliography}
\end{document}